\documentclass[aps,prd,showkeys,nofootinbib,amsmath,amssymb,superscriptaddress,onecolumn]{revtex4-2}
\usepackage{booktabs}
\usepackage{dsfont}
\usepackage{mathtools} 
\usepackage{xcolor}
\definecolor{nicegreen}{rgb}{0.07, 0.564, 0.04}
\usepackage{framed}
\usepackage{comment}
\usepackage{xspace}
\usepackage{graphicx}% Include figure files
\usepackage{dcolumn}% Align table columns on decimal point
\usepackage{bm}% bold math
\usepackage{tikz-cd}
\usepackage[normalem]{ulem}
\usepackage{slashed}

\newcommand\ie{\mbox{\textit{i.\,e.}}\xspace}

\newcommand\eg{\mbox{e.\,g.}\xspace}

\newcommand\D{\mathrm{d}}
\newcommand{\hx}{\hat{x}}
\newcommand{\hp}{\hat{p}}
\newcommand{\hk}{\hat{k}}

\newcommand{\hP}{\hat{P}}
\newcommand{\mompar}{\dot{\partial}}

\DeclarePairedDelimiter\braket{\langle}{\rangle}
\DeclarePairedDelimiterX\Braket[2]{\langle}{\rangle}{#1 \delimsize\vert #2}

\begin{document}

\title{The minimal length: a cut-off in disguise?}

\author{Pasquale Bosso}\email[]{pbosso@unisa.it}
\affiliation{Dipartimento di Ingegneria Industriale, Universit\`a degli Studi di Salerno, Via Giovanni Paolo II, 132 I-84084 Fisciano (SA), Italy}

\author{Luciano Petruzziello}\email[]{lupetruzziello@unisa.it}
\affiliation{Dipartimento di Ingegneria Industriale, Universit\`a degli Studi di Salerno, Via Giovanni Paolo II, 132 I-84084 Fisciano (SA), Italy}
\affiliation{INFN, Sezione di Napoli, Gruppo collegato di Salerno, Via Giovanni Paolo II, 132 I-84084 Fisciano (SA), Italy \\
Institut f\"ur Theoretische Physik, Albert-Einstein-Allee 11, Universit\"at Ulm, 89069 Ulm, Germany}

\author{Fabian Wagner}
\email[]{fwagner@unisa.it}
\affiliation{Dipartimento di Ingegneria Industriale, Universit\`a degli Studi di Salerno, Via Giovanni Paolo II, 132 I-84084 Fisciano (SA), Italy}
\begin{abstract}
    The minimal-length paradigm, a possible implication of quantum gravity at low energies, is commonly understood as a phenomenological modification of Heisenberg's uncertainty relation. We show that this modification is equivalent to a cut-off in the space conjugate to the position representation, \ie the space of wave numbers, which does not necessarily correspond to momentum space. This result is generalized to several dimensions and noncommutative geometries once a suitable definition of the wave number is provided.
    Furthermore, we find a direct relation between the ensuing bound in wave-number space and the minimal-length scale. 
    For scenarios in which the existence of the minimal length cannot be explicitly verified, the proposed framework can be used to clarify the situation. Indeed, applying it to common models, we find that one of them 
    does, against all expectations, allow for arbitrary precision in position measurements. In closing, we comment on general implications of our findings for the field. In particular, we point out that the minimal length is purely kinematical such that, effectively, there is only one model of minimal-length quantum mechanics.
\end{abstract}

\maketitle

When regularizing in quantum field theory, it is often (if somewhat na\"ively) concluded that a finite cut-off in relativistic momentum space regularizing UV-divergences implies the existence of an underlying lattice structure. The corresponding lattice spacing provides a minimal length. In the literature on conventional minimal-length theories, on the other hand, it is common to interpret the minimal-length scale not as a physical length, but as a limit to the physically attainable resolution in distance measurements \cite{Mead:1964zz,Garay:1994en,Kempf:1994su,Scardigli:1999jh,Hossenfelder:2012jw,Wagner:2022bcy}. In quantum mechanics, for example, this corresponds to a minimum for the standard deviation of the position operator
\begin{equation}
    \Delta x_a\geq\ell,\label{MinimumLength}
\end{equation}
with the newly introduced length scale $\ell.$
This interpretation attributes a fundamental ``fuzziness'' to the background spacetime itself owing to a modification of the Heisenberg algebra. Notwithstanding the apparent difference from the conventional cut-off, following \cite{Kempf:1996nk} this kind of assumption has been used frequently to regularize integrals in phenomenological applications such as the brick wall model of black hole thermodynamics \cite{Liu:2003ke,Liu:2005iwa,Zhao:2006xf}. One may thus wonder in which way the minimal-length idea differs from a physical cut-off in momentum space.

In this paper, we show that a minimal-length scale as given in \eqref{MinimumLength} is indeed equivalent to a hard cut-off. However, this cut-off is not bounding momentum space, but rather the space of wave numbers which we define as being the space conjugate to the position representation. As a matter of fact, it is possible to explicitly relate the bound in wave-number space to the minimal-length scale. Yet, a deformation of the Heisenberg algebra immediately implies a modification of the de Broglie relation such that wave numbers and momenta cease to be proportional to each other \cite{Hossenfelder:2012jw,Bosso:2020aqm}. Therefore, momentum space may be unbounded despite wave-number space is not. This, it turns out, is the subtle difference in interpretation between applying a cut-off and deforming the Heisenberg algebra. Bear in mind, however, that the definition of a ``physical'' momentum cannot be motivated from the minimal length itself.

The interpretational difference becomes all the more pronounced once the coordinates become noncommutative. To cover this possibility, we generalize the concept of wave number to deformed Heisenberg algebras which entail a noncommutative geometry. The resulting (anisotropic) wave-number space continues to be bounded under the assumption of a minimal length. Similarly, the relation between this bound and the minimal-length scale can be generalized.

The present approach is far from being only of conceptual interest. It can be used as a tool to identify deformed Heisenberg algebras which possess a minimal localization and those that do not -- also in situations where this may not be possible by other means. Applying this reasoning to the most commonly used models of the field, we indeed find one which, contrary to claims in the literature \cite{Maggiore:1993kv,Fadel:2021hnx}, does not encompass a minimal-length scale. 

To comply with the above purposes, the paper is structured as follows: first, we propose the argument for a bound in wave-number space for one spatial dimension (or equivalently for multiple commutative dimensions) in Section \ref{sec:com}. This result is then generalized to noncommutative geometries in Section \ref{sec:noncom}. Section \ref{sec:application} is devoted to the application of the general framework to existing models. In section \ref{sec:essence} we comment on the general implications of our results for minimal-length models. Finally, we summarize and discuss our findings in Section \ref{sec:discussion}.

Throughout the work we will use natural units $\hbar=c=1$.

\section{No cut-off, no minimal length}\label{sec:com}

Let us first consider one spatial dimension, and assume the position of the system at hand to be given by the operator $\hat{x}.$ Then, we may always find a conjugate wave number operator $\hat{k}$ such that the ordinary Heisenberg algebra
\begin{equation}
    [\hx,\hk]=i,\label{HeisAlg}
\end{equation}
is satisfied.
While $\hk$ is not regarded as the physical momentum operator in conventional minimal-length theories, it is bound to exist, and can be used to construct a representation of the underlying deformed Heisenberg algebra.

In this Section, we will show that a lower bound of type \eqref{MinimumLength} is equivalent to a bounded spectrum for $\hk$.
This means that the minimal-length constraint imposes a cut-off on the conjugate wave-number space.
Intuitively, one would expect that to happen: given a pair of observables satisfying the Heisenberg algebra \eqref{HeisAlg}, if the spectrum of $\hk$ is continuous and unbounded, it is a simple exercise to construct states which violate the inequality \eqref{MinimumLength}.

Consider, thus, a quantum system confined to a box of length $2B$ in wave-number space, \ie $\text{spec}(\hk)=\{k: k\in [-B,B]\}.$ To achieve this, we apply Dirichlet boundary conditions at $k=\pm B$.
Clearly, we can express any state $\psi$ in terms of the eigenstates of $\hat{x}^2$ as
\begin{equation}
    \psi=\sum_{n=0}^\infty \left[a_n\frac{\sin [(n+1)\pi k/B]}{\sqrt{B}}+ b_n \frac{\cos\left[(2n+1)\pi k/2B\right]}{\sqrt{B}}\right],
\end{equation}
with the complex coefficients $a_n,$ $b_n$ satisfying $\sum_{n=0}^\infty(|a_n|^2+|b_n|^2)=1.$ 
Since $\hx$ obeys the Heisenberg algebra with $\hk$, it can be represented as a derivative with respect to $k$, \ie $\hx\psi=i\partial_k\psi.$ We thus obtain for a generic state
\begin{align}
    \Delta x^2\equiv&\braket{\psi|\hx^2\psi}-\braket{\psi|\hx\psi}^2\leq\braket{\hx^2}=\left(\frac{\pi}{2B}\right)^2\sum_{n=0}^\infty \left[4|a_n|^2(1+n)^2+|b_n|^2(1+2n)^2\right].
\end{align}
The right-hand-side of this inequality is clearly minimal if $|b_0|=1$ while all other coefficients vanish.
Assuming a minimal length of the kind \eqref{MinimumLength}, we then obtain
\begin{equation}
    \ell^2\leq\Delta x^2\leq \left(\frac{\pi}{2B}\right)^2.\label{MinLengthCutOff}
\end{equation}
Standard quantum mechanics would be recovered in the limit $B\rightarrow \infty.$ However, this would violate inequality \eqref{MinLengthCutOff}, \ie it is impossible in the presence of a minimal length.
Hence, a theory characterized by a minimal length cannot be described in terms of an unbounded wave number.

Starting from the above premises, the argument can be refined even more:
it is possible to relate the bound
$B$ of the wave-number spectrum to the minimal length $\ell$.
To this aim, we first notice that the above model does not yield a preferred position.
Therefore, there have to be states of smallest possible position uncertainty for every $\braket{\hx},$ all of which produce the same value for $\Delta x.$ Then, it is sufficient to consider states satisfying $\braket{\hx}=0.$ Under these circumstances, the smallest possible position uncertainty is indeed given by
\begin{equation}
    \Delta x=\frac{\pi}{2B}.
\end{equation}
This quantity is bounded by the minimal length, thereby leading to the fundamental bound in wave-number space
%which provides the closest possible approximation to the unbounded case
\begin{equation}
    B=\frac{\pi}{2\ell}.\label{1Dbound}
\end{equation}
Thence, provided that there is a minimal length for the position operator, the spectrum of the corresponding conjugate wave number operator is
\begin{equation}
    \text{spec}(\hk)=\{k:k\in [-\pi/2\ell,\pi/2\ell]\}.\label{WaveNumberSpec}
\end{equation}
In turn, a quantum theory in which the wave number conjugate to the position does not have a bounded spectrum does not have a minimal length. This result can be straightforwardly generalized to several spatial dimensions as long as the underlying geometry is commutative. As we will see in the following Section, noncommutative geometries are slightly more involved to deal with. 

\section{Noncommutative geometry}\label{sec:noncom}

In general, minimal-length models are not understood in terms of conjugate variables $\hx$ and $\hk.$ Instead, they are based on a modified Heisenberg algebra expressed with the help of a ``physical'' momentum $\hP_a,$ say 
\begin{align}
    [\hx_a,\hP_b]=i\left[f(\hP^2)\delta_{ab}+\Bar{f}(\hP^2)\frac{\hP_a\hP_b}{\hP^2}\right],&&[\hP_a,\hP_b]=0,\label{DefHeisAlg1}
\end{align}
with the two functions $f,$ $\Bar{f}$ constrained to reduce to $1$ and $0$ in the low-energy limit, that is $P^2\rightarrow 0$, so as to guarantee the recovery of the Heisenberg algebra.
Unless these two functions satisfy the relation \cite{Wagner:2021bqz,Bosso:2022ogb}
\begin{equation}
    \bar{f}=\frac{2(\log f)'\hP^2}{1-2(\log f)'\hP^2}f,
\end{equation}
the coordinates $\hx_a$ of the model \eqref{DefHeisAlg1} fail to commute. In this scenario, one can verify that
\begin{equation}
    [\hx_a,\hx_b]\propto 2\hx_{[b}\hP_{a]},
\end{equation}
where the proportionality factor depends on $\hP^2,$ and is related to the functions $f$ and $\bar f$ via Jacobi identities. If coordinates are noncommutative in this way, there is no possibility to recover the undeformed $d$-dimensional Heisenberg algebra by merely choosing a new wave number-like variable while keeping the coordinates $\hx_a$ as they are; clearly, if we continue to use the $\hx_a,$ the noncommutativity cannot be forced to disappear.
Nevertheless, we can find a set of wave numbers which are conjugate to the respective coordinates. 

In order to achieve this, it is instructive to follow a two-step procedure: firstly, we diagonalize the deformed Heisenberg algebra; secondly, we find a transformation which restores the undeformed Heisenberg algebra on the diagonal.

In that vein, we define another momentum coordinate $\hp_a=\bar{g}(\hP^2)\hP_a.$ After some algebra, it can be shown that
\begin{equation}
    [\hx_a,\hp_b] = if\Bar{g}\delta_{ab}+i\left[\bar{g}\Bar{f}+2\left(f+\Bar{f}\right)\Bar{g}'\hP^2\right]\frac{\hP_a\hP_b}{\hP^2}.
\end{equation}
Here, we choose the second term of this equation to vanish.
Accordingly, $\bar{g}$ assumes the form
\begin{equation}
    \bar{g}=\exp\left(-\int_0^{\hP^2}\frac{\Bar{f}(\Pi)}{2\Pi\left[f(\Pi)+\Bar{f}(\Pi)\right]}\D\Pi\right),
\end{equation}
where as usual $\bar{g}(0)=1,$ implying that the momenta $\hp_a$ and $\hP_a$ are equal in the low-energy limit.
As a result, we obtain the diagonal deformed Heisenberg algebra
\begin{equation}
    [\hx_a,\hp_b]=i\delta_{ab} ~ g\circ \hP^2(\hp^2),
    \label{eqn:diagonal}
\end{equation}
where we defined $g\equiv f\Bar{g}.$ For the sake of conciseness, henceforth we will omit the composition with $\hP^2$. Next, imposing the Jacobi identities, one can check that a diagonal algebra of this kind implies a commutator of the coordinates given by
\begin{equation}
    [\hx_a,\hx_b]=2g' \hx_{[b}\hp_{a]}\equiv \theta\hx_{[b}\hp_{a]},
\end{equation}
%where the last equality defines the noncommutativity $\theta(\hp^2).$
where we introduced the noncommutativity $\theta(\hp^2) = 2 g'(\hp^2)$. We note that, in the case of a commutative geometry, this first step would have already led to the Heisenberg algebra, \ie $\theta=g'=0$ would immediately imply $g=1$.
As a result, we could directly employ the reasoning laid out above for the one-dimensional case to conclude that the space spanned by the $\hp_a$ is to be bounded for a minimal length to appear.

For noncommutative geometries, however, we have to resort to a second transformation. To better convey the reason behind this step, it is instructive to consider the one-dimensional counterpart of the algebra \eqref{eqn:diagonal}. In this case, the wave number is related to the momentum $\hp$ as \cite{Bosso:2022vlz}
\begin{equation}
    \hk=\int_0^{\hp}\frac{\D p'}{g(p^{\prime 2})}.\label{eqn:GenMomWaveNum}
\end{equation}
In several dimensions, we can introduce the analogous transformation
\begin{equation}
    \hk_a=\int_0^{\hp_a}\frac{\D p'_a}{g\left(p_a^{\prime 2}+\sum_{b\neq a}\hp_b^2\right)},\label{WaveNumberNonComGeo}
\end{equation}
where we have explicitly separated the dependence on the component $\hp_a$ from the other components, with $b\neq a$, in the function $g(p^2)$.
This transformation is particularly nontrivial, because it does not preserve the isotropy of the underlying uncertainty relations:
the integration is performed along a specific axis in momentum space, which introduces a preferred coordinate system.
Therefore, a rotation in wave-number space does not correspond to a rotation of positions or momenta. This can be seen from the Jacobian
\begin{align}\label{Jacobian}
    J_{ab}=\frac{\partial \hk_a}{\partial \hp_b}=\begin{cases}
        g^{-1}(\hp^2)&a=b,\\
        -\hp_b\int_0^{\hp_a}\frac{\theta\D p'_a}{g^2}&a\neq b,
    \end{cases}
\end{align}
which is nontrivial if the noncommutativity is different from 0, and cannot be expressed covariantly.
Despite this, we still have $\hat{p}_a|_{k_a=0}=0$ such that $J_{ab}|_{k_a=0}=J_{ab}|_{k_b=0}=\delta_{ab}/g.$ Consequently, the algebra of observables becomes
\begin{align}
    [\hx_a,\hk_b]= i gJ_{ba}=  \begin{cases}
                         i & a=b,\\
                         -ig\hp_a\int_0^{\hp_b}\frac{\theta}{g^2}\D p'_b &a\neq b.
                    \end{cases}
                    \label{eqn:comm_xk}
\end{align}
On the diagonal, the positions and the effective wave numbers satisfy the one-dimensional Heisenberg algebra (\ie the wave numbers are conjugate to the respective coordinates). In general, the wave-number spectrum may be bounded to some domain $D$ which, due to the anisotropic nature of the transformation \eqref{WaveNumberNonComGeo}, may not be isotropic. We will see some examples of this in Section \ref{sec:application}.
The anisotropies crucially depend on the noncommutativity $\theta$ of the coordinates and vanish for commutative backgrounds.

In light of the above, a question naturally arises: does the minimal length still imply a cut-off in this wave-number space? More precisely, is the lowest eigenvalue of the squared position in a given direction, say $(\hat{x}_d)^2,$ related to such a bound? 
We answer both questions in a representation-independent fashion in Appendix \ref{apx:min_len_non_comm}. Here, we provide a simplified argument.

First, let us make an observation: if the background possesses non-vanishing spatial non-commutativity $\theta,$ the coordinates satisfy the uncertainty relation
\begin{equation}
    \Delta x_a\Delta x_b\geq \frac{1}{2} |\langle\theta (\hp^2) \hp_{[a}\hx_{b]}\rangle|.\label{CoordinateUncertainty}
\end{equation}
To minimize the uncertainty along the direction $x_d$, we need to consider states with large uncertainties in all orthogonal directions. In other words, we require $\Delta x_b\to\infty$ for all $b\neq d$, thus demanding that a state characterized by the smallest uncertainty $\Delta x_d$ be infinitely peaked in momentum space in those directions.
By virtue of Eq. \eqref{WaveNumberNonComGeo}, the property of being peaked in the origin carries over to the wave numbers.
To further minimize the effect of the non-commutativity, whose absolute value (at least around the origin in momentum space) increases monotonically with $\hp^2$, it is to be expected that the peak should be situated in the origin of the respective directions. Indeed, for those states infinitely peaked in the origin, it can be shown that the right-hand-side of Eq. \eqref{CoordinateUncertainty} always vanishes, \ie they are not affected by the coordinate noncommutativity. This way, we can study the minimal length independently of the influence of the noncommutativity. 

Furthermore, as effects of the geometry cease to play a r\^ole, the wave function saturates the uncertainty relations involving positions or wave numbers in the directions normal to $p_d$ (this can also be inferred from the Jacobian \eqref{Jacobian} being diagonal at vanishing involved wave numbers). Investigating the states saturating uncertainty relations, in turn, is equivalent to investigating the underlying uncertainty relations themselves.

In momentum space, such a projection on the $d-$th axis can be obtained by reducing the state space to wave functions
\begin{equation}
    \psi\simeq\psi_d(p_d)\prod_{j=1}^{d-1}\frac{e^{-\frac{p_j^2}{4\epsilon}}}{\sqrt[4]{2\pi\epsilon}}.\label{ApproxTestFunction}
\end{equation}
In the end, we will take the limit $\epsilon\rightarrow 0,$ thereby imposing that the involved Gaussians are infinitely peaked in the origin of momentum space. In the following, we intend to evaluate the position uncertainty in the $d-$th direction given the states \eqref{ApproxTestFunction}.

As every modified Heisenberg algebra can be reduced to the diagonal type \eqref{eqn:diagonal} by mere redefinition of momenta, we assume it to be the starting point. As a result, we may consider the momentum representation of the position operator
\begin{equation}
    \hx_a\psi=ig(p^2)\mompar_a\psi,\label{PosOpMomSpace}
\end{equation}
where we introduced the momentum derivative $\mompar_a=\partial/\partial p_a.$ The position operator is symmetric with respect to the integration measure $\D^dp/g.$ Without loss of generality, we consider states with vanishing expected position $\braket{x_d}$ (again there is no preferred position in the model). Therefore, we can write
\begin{align}
    \Delta x_d^2=&-\int_{\mathcal{D}_p}\frac{\D^dp}{g(p^2)}\psi^*\left[g(p^2)\mompar_d\right]^2\psi=-\int_{\mathcal{D}_p}\frac{\psi_d^*\left[g(p^2)\mompar_d\right]^2\psi_d}{g(p^2)}\left(\prod_{j=1}^{d-1}\frac{e^{-\frac{p_j^2}{2\epsilon}}}{\sqrt{2\pi\epsilon}}\D p_j\right)\D p_d,
\end{align}
where the domain of integration $\mathcal{D}_p$ depends on the choice of the model.
For vanishing $\epsilon$, the product in brackets just becomes a product of Dirac delta-distributions 
\begin{align}
    \lim_{\epsilon\to 0}\Delta x_d^2=\int_{\mathcal{D}_p}\frac{\psi_d^*\left[g(p^2)\mompar_d\right]^2\psi_d}{g(p^2)}\left(\prod_{j=1}^{d-1}\delta (p_b)\D p_b\right)\D p_d,
\end{align}
where we applied the definition of the Dirac delta-distribution as an infinitely peaked Gaussian, \ie $\delta (x)=\lim_{\epsilon\to 0}e^{-x^2/2\epsilon}/\sqrt{2\pi\epsilon}.$ Consequently, the integration is trivial, and we may immediately project on the $d-$th dimension to obtain
\begin{equation}
    \lim_{\epsilon\to 0}\Delta x_d^2=-\int_{-\Bar{p}_d}^{\Bar{p}_d}\frac{\D p_d}{g(p_d^2)}\psi_d^*\left[g(p_d^2)\mompar_d\right]^2\psi_d,
\end{equation}
where, according to the model at hand, the effective bound to momentum space in the $d-$th dimension $\bar p_d$ may be finite or infinite. At this point, redefining the integration variable as $\D \bar{k}_d=\D p_d/g(p_d^2)=\D k|_{p_b=0}$ for all $b\neq d$ (which is indeed just the transformation \eqref{WaveNumberNonComGeo} for vanishing transverse momenta), we obtain
\begin{equation}
    \lim_{\epsilon\to 0}\Delta x_d^2=-\left.\int_{-B}^B\D k_d\psi_d^*\frac{\partial^2\psi_d}{\partial k_d^2}\right|_{p_b=0},
\end{equation}
%with $b\neq d$ and 
where, similarly to the one-dimensional case, $B$ may be finite or infinite.
The effective one-dimensional operator $\hx_d\psi = i\partial/\partial k_d\psi_d|_{p_b=0}$ is clearly unmodified with respect to the case of commuting coordinates. In short, for vanishing spread of the wave function $\psi$ (given in Eq. \eqref{ApproxTestFunction}) in the transverse directions of momentum space (the limit $\epsilon\rightarrow 0$), the position uncertainty in the longitudinal direction is not affected by the presence of coordinate noncommutativity.

Hence, recalling the argument outlined in Section \ref{sec:com}, if $B$ is infinite the position uncertainty can be made arbitrarily small.
If it is not, the effective value of the bound can be related to the minimal length as 
\begin{equation}
    B=\lim_{\hp_d\to \Bar{p}_d}\left(\prod_{b=1}^{d-1}\lim_{\hp_b\to 0}\right)\hk_d=\frac{\pi}{2\ell}.\label{BoundSeveralD}
\end{equation}
In a nutshell, the fact that a minimal length requires a bounded wave-number space holds true also for noncommutative scenarios if we define the wave numbers by the transformation \eqref{WaveNumberNonComGeo}.

Having shown that the approach of the present paper is valid also in several, possibly noncommutative dimensions, we are ready to apply it to existing models in the literature in order to check for the existence of minimal-length scales.

\section{To bound or not to bound}\label{sec:application}

Given a model in the shape \eqref{eqn:diagonal}\footnote{As shown in the previous Section, every model can be cast in this form by simple redefinition of the momentum operator.}, we have shown that the domain of the wave number defined in Eq. \eqref{WaveNumberNonComGeo} has to be bounded for the model to have a minimal length. This is especially the case in the limit of vanishing transverse wave numbers as seen in Eq. \eqref{BoundSeveralD}, which is in complete correspondence to Eq. \eqref{MinLengthCutOff}. 

Let us first consider the one-dimensional counterpart of the model \eqref{eqn:diagonal}, namely
\begin{equation}
    [\hx,\hp]=ig(\hp^2).
\end{equation}
This algebra can be brought into a canonical form by finding the corresponding $\hk (\hp)$ which is conjugate to the $\hx,$ \ie finding Darboux-coordinates without modifying $\hx.$ This has already been done in all generality in Eq. \eqref{eqn:GenMomWaveNum}. By virtue of Eq. \eqref{WaveNumberSpec}, whether the model at hand possesses a minimal length depends on the image of the function $\hk (\hp)$ being bounded for allowed values of $\hp$.\footnote{Some models also predict a maximal momentum $\hp,$ \ie a bounded momentum space. This can be read off from the preimage within which $\hk(\hp)$ is an invertible map, meaning that the Jacobian (see Eq. \eqref{Jacobian}) is non-degenerate.} Thus, we can immediately obtain the exact value of the minimal length.

A short inspection of Eq. \eqref{eqn:GenMomWaveNum} shows that it is equivalent to Eq. \eqref{WaveNumberNonComGeo}, say in direction $d,$ at vanishing transverse momenta
\begin{equation}
    \left(\prod_{a\neq d}\lim_{p_a\to 0}\right)\hk_d=\int_0^{\hp_d} \frac{\D\hp'_d}{g(\hp_d^{\prime 2})}.\label{eqn:WaveNumberNonComGeoVanTrans}
\end{equation}
Domain and image of Eqs. \eqref{eqn:WaveNumberNonComGeoVanTrans} and \eqref{eqn:GenMomWaveNum}, and with them the respective bounds \eqref{BoundSeveralD} and \eqref{1Dbound}, are clearly the same. Therefore, it is sufficient to consider all models in one dimension to search for the minimal length.

\subsection{One class of common models}

Typically, the majority of the models investigated in the literature on deformed Heisenberg algebras \cite{Kempf:1994su,Kempf:1996fz,Maggiore:1993kv,Fadel:2021hnx} belongs to one class, which is characterized by a relation of the form 
\begin{equation}
    [\hat{x},\hat{p}] = i \hbar (1 + \beta \hat{p}^2)^\alpha,\label{eqn:modelclass}
\end{equation}
where $\alpha>0$ identifies the model at hand while $\beta,$ having units of $[l^{2}],$ provides a length scale. This length scale is commonly associated with the minimal length. However, it is only in the case $\alpha=1,$ yielding $\ell=\sqrt{\beta}$ (see \cite{Kempf:1994su}), that this connection can be worked out explicitly by applying the Robertson-Schr\"odinger relation \cite{Schroedinger:1930awq,Robertson:1929zz}.\footnote{While in \cite{Fadel:2021hnx} it has been claimed to have been shown for the case $\alpha=1/2$ as well, there is a flawed step between Eqs. (23) and (24) in that reference, explaining the divergence of this conclusion from our results below.}

\begin{figure}
    \centering
    \includegraphics{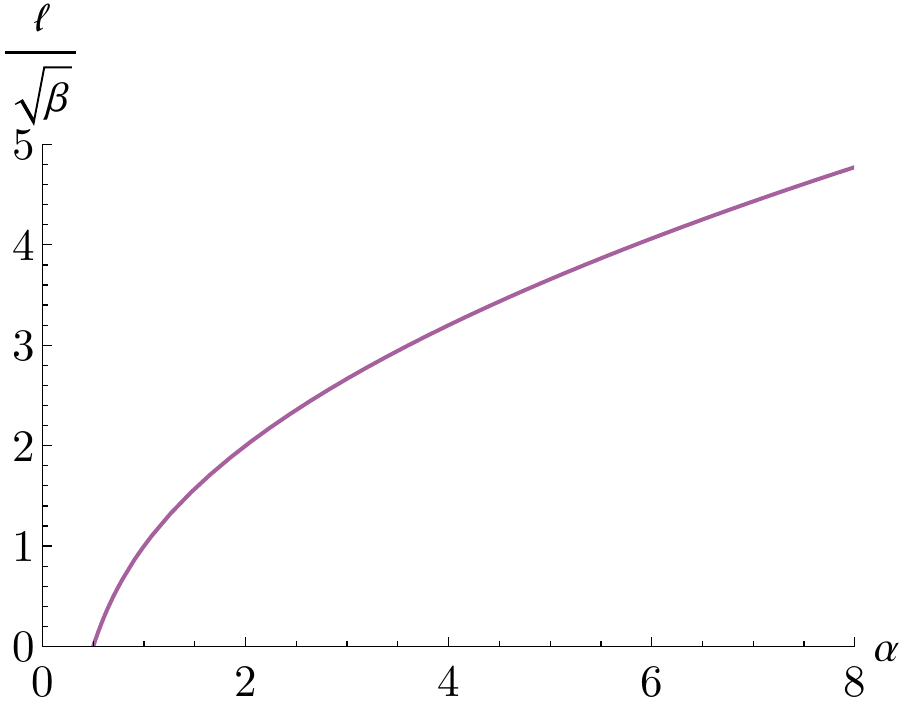}
    \caption{Value of the minimal length for the one-parameter family of minimal-length models \eqref{eqn:modelclass} as a function of the model classifier $\alpha$ evaluated in terms of the model parameter $\beta.$}
    \label{fig:AlphaModels}
\end{figure}

This is where the strength of the present approach comes in. Given a model of the kind \eqref{eqn:modelclass}, all we have to do is finding the wave number $\hk,$  and investigating its domain.
By resorting to Eq. \eqref{eqn:GenMomWaveNum}, the wave number and momentum operators are related by the expression \cite{Kempf:1994su,Bosso:2020aqm}
\begin{align}
    \hk(\hp)
    = \int_0^{\hp} \frac{\D\hp'}{(1+\beta \hp^{\prime 2})^\alpha}
    = \frac{\hp}{\sqrt{1 + \beta p^2}} ~ {}_2F_1 \left(\frac{1}{2}, \frac{3}{2}-\alpha;\frac{3}{2};\frac{\beta \hp^2}{1 + \beta \hp^2}\right),
    \label{eqn:alphamodelsint}
\end{align}
where ${}_2F_1$ is the Gaussian hypergeometric function.
To evaluate the limit $\hp \to \infty$ for any positive value of $\alpha$, it is convenient to differentiate the models with $\alpha\leq 1/2$ from the ones where $\alpha>1/2.$
\begin{itemize}
    \item $\alpha \leq \frac{1}{2}:$ for these models we find
\begin{equation}
    \hk(\hp) 
    \geq \int_0^{\hp} \frac{\D\hp'}{\sqrt{1+\beta \hp^{\prime 2}}} 
    = \frac{\text{arcsinh}\left(\sqrt{\beta } \hp\right)}{\sqrt{\beta }}.
    \label{eqn:k_l1/2}
\end{equation}
Both image and domain of this function are unbounded. In other words, $\hp$-space is unbounded and $\hk$ diverges in the limit $\hp\to\infty$. Thus, these models do not incorporate a minimal length.
\item $\alpha> 1/2:$ in this case, using Gauss' summation theorem \cite{bailey1964generalized}, we have
\begin{equation}
    \lim_{p \to \infty} \frac{p}{\sqrt{1 + \beta p^2}} ~ {}_2F_1 \left(\frac{1}{2}, \frac{3}{2}-\alpha;\frac{3}{2};\frac{\beta p^2}{1 + \beta p^2}\right)
    = \frac{\sqrt{\pi} \Gamma(\alpha - \frac{1}{2})}{2 \sqrt{\beta} \Gamma(\alpha)} 
\end{equation}
which is finite for $\alpha>\frac{1}{2},$ implying the minimal length
\begin{equation}
    \ell=\frac{\sqrt{\pi\beta}\Gamma (\alpha)}{\Gamma (\alpha-\frac{1}{2})}.
\end{equation}
This function is displayed in Fig. \ref{fig:AlphaModels}. As can be gathered from there, the minimal length decreases for decreasing $\alpha$ and vanishes at the boundary value $\alpha=1/2.$ Furthermore, for $\alpha=1,$ we obtain $\ell=\sqrt{\beta}$, in exact correspondence with the result derived from the Robertson-Schr\"odinger relation \cite{Kempf:1994su}. 
\end{itemize}

\begin{figure}
    \centering
    \includegraphics{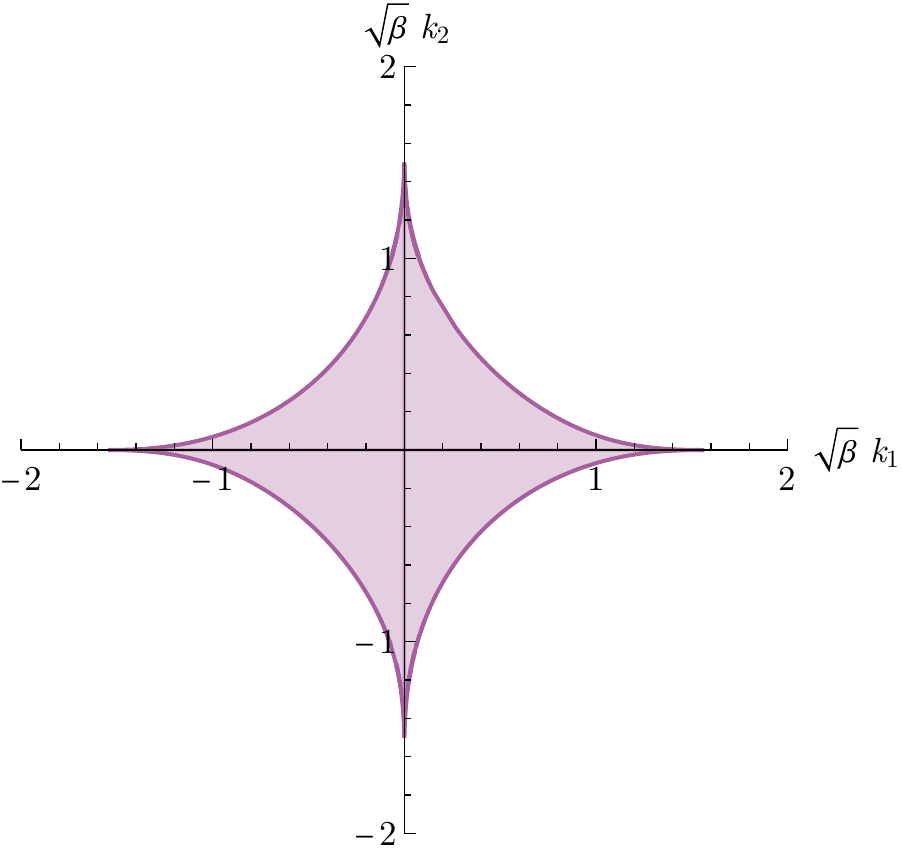}
    \caption{Domain of the wave number $\hk$ for the case $\alpha=1$ of the family of models in Eq. \eqref{eqn:modelclass} in two dimensions. The region is bounded. In particular, on the axes the wave numbers do not exceed the value $\pi/2\sqrt{\beta}.$}
    \label{fig:my_labelkmm}
\end{figure}

To show how the case $\alpha=1$ \cite{Kempf:1994su} plays out in two dimensions, the region of allowed wave numbers is displayed in Fig. \ref{fig:my_labelkmm}. It is clearly bounded. In particular, at vanishing transverse wave number, \ie on the axes, the bound equals exactly $\pi/2\sqrt{\beta}$ as expected. Furthermore, it is possible to see the anisotropy of the wave-number representation reflected in the star-like shape of the region.

\begin{figure}
    \centering
    \includegraphics{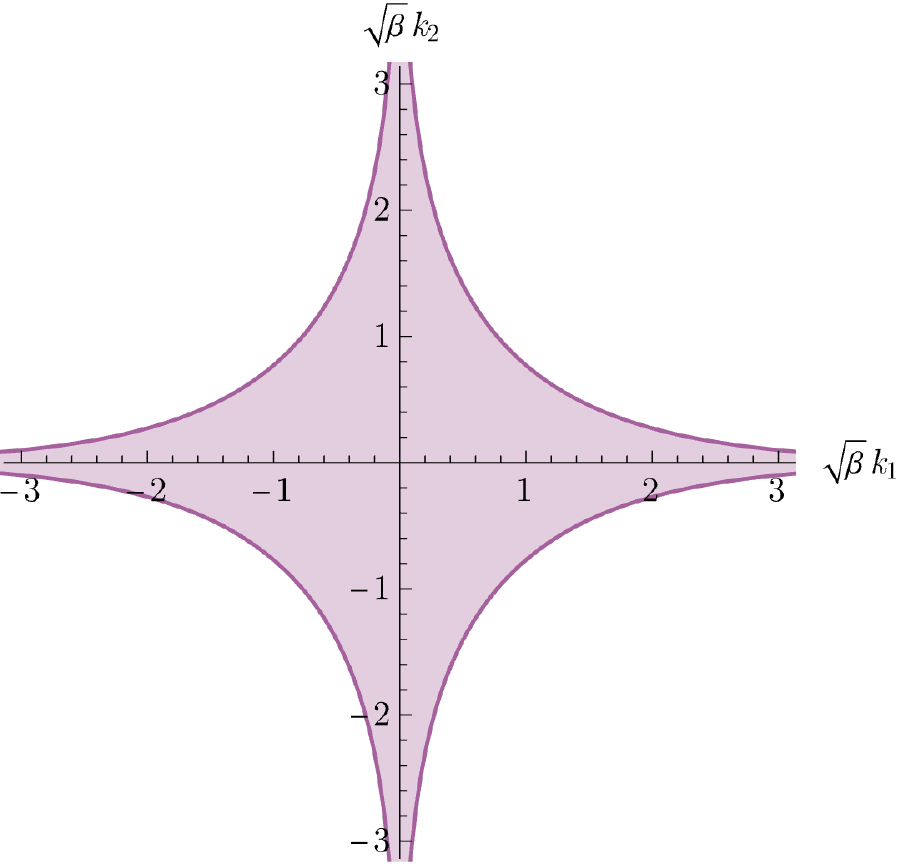}
    \caption{Domain of the wave number $\hk$ for the case $\alpha=1/2$ of the family of models in Eq. \eqref{eqn:modelclass} in two dimensions.
        Notice that such a model is characterized by an unbounded domain.
        Specifically, when $k_1 = 0$, $k_2$ can acquire any real value, and vice-versa.}
    \label{fig:my_labelmm}
\end{figure}

The boundary case $\alpha=1/2$ is of particular interest due to it having been the basis for one of the very foundational works of the field \cite{Maggiore:1993kv}.
We show the domain of its wave-number space in two dimensions in Fig. \ref{fig:my_labelmm}. In contrast to the example $\alpha=1$, in Fig. \ref{fig:my_labelkmm} this region is clearly unbounded.
To support our finding of this model not possessing a minimal length, we have explicitly constructed states which satisfy the proposed uncertainty relation and at the same time allow for infinite localizability in Appendix \ref{app:maggiore}.

\subsection{Other models}

There are a number of other common ans\"atze which are not of the kind \eqref{eqn:modelclass}.
These models may even have a bounded momentum ($\hp$) space %, \ie $\hp$-representation, 
but no minimal length or vice-versa. The results for some of them are summarised in Table \ref{tab:DifferentModels}. We find that, contrary to the claim in \cite{petr} by one of the authors of the present paper, the model $g=\sqrt{1-\beta\hp^2}$ does actually predict a minimal length. Apart from that, the bounds reflect what was known in the literature.

\bgroup
\def\arraystretch{1.55}
\begin{table}[h]
    \centering
    \begin{tabular}{@{\hspace{1em}}c@{\hspace{1em}}|@{\hspace{1em}}c@{\hspace{1em}}|@{\hspace{1em}}c@{\hspace{1em}}|@{\hspace{1em}}c@{\hspace{1em}}|c}
        \toprule
        $g(\hp^2)$ & wave number $\hk (\hp)$ & maximal momentum ($\hp$) & minimal length & Ref. \\
        \midrule
         %$1$ & $\hp$ & none & none\\
         $1-\beta \hp^2$& $\text{arctanh}(\sqrt{\beta}\hp)/\sqrt{\beta}$ & $1/\sqrt{\beta}$ & none & \cite{Jizba:2009qf,Ong:2018zqn}\\
         $e^{\beta \hp^2}$ & $\frac{\sqrt{\pi}}{2\sqrt{\beta}}\text{Erf}(\beta\hp)$ & none & $\sqrt{\pi\beta}$ & \cite{Nouicer:2007jg}\\
         $\frac{1}{1-\beta\hp^2}$ & $\hp\left(1-\frac{\beta\hp^2}{3}\right)$ & $1/\sqrt{\beta}$ & $3\pi\sqrt{\beta}/4$ &\cite{Pedram:2011gw,Pedram:2012my}\\
         $\sqrt{1-\beta\hp^2}$ & $\frac{\text{arcsin}(\sqrt{\beta}\hp)}{\sqrt{\beta}}$ & $1/\sqrt{\beta}$ & $\sqrt{\beta}$ & \cite{petr}\\
         \bottomrule
    \end{tabular}
    \caption{Wave numbers, momentum-space bounds (if existent) and minimal lengths (if existent) for common deformed Heisenberg algebras. The last column indicates the references associated to the models.}
    \label{tab:DifferentModels}
\end{table}
\egroup

Having gathered the results on different minimal-length models, it is time to comment on the minimal length and how the multitude of distinct realizations of it is to be interpreted.

\section{The essence of the minimal length}\label{sec:essence}

Throughout this paper, we have aimed at distilling the very foundation of the minimal-length idea. Nevertheless, we have never had to refer to the dynamics of a system, \ie its Hamiltonian. This explicitly shows that the minimal length is to be explained on the level of kinematics. It is a property of the background on top of which we define a quantum theory. 

Any interpretation in terms of a ``physical'' momentum (throughout the paper denoted as $\hp$ or $\hp_a$) which satisfies some modified Heisenberg algebra requires additional structure, while the bound in wave-number space is sufficient to fully characterize the minimal length. The ad-hoc definition of an additional momentum (which may indeed be useful from the point of view of interpretation or calculation) has no physical consequences. However, the choice of Hamiltonian made in the foundational papers on minimal-length quantum mechanics (\eg \cite{Kempf:1994su})
\begin{equation}
    H=\frac{\hp^2}{2m}+V(x),\label{eqn:GUPHam}
\end{equation}
and countless times in the literature since, does inherit a degree of arbitrariness from it.
Why, for example, should we not choose the Hamiltonian
\begin{equation}
    H=\frac{\hk^2}{2m}+V(x)\label{UnpertWaveHam}
\end{equation}
instead, as suggested in \cite{Bosso:2022rue}?
The effect of the minimal length would still be included by the bound in wave-number space.
We thus see that all different minimal-length models, while being kinematically equivalent, only differ in their dynamics, and there is no physical reason to prefer one model over the other (as long as both actually predict a minimal length). In other words, the multitude of approaches only add a layer of modification to the Hamiltonian, which cannot be motivated by the existence of a minimal length itself.\footnote{In the context of noncommutative backgrounds -- themselves an additional assumption -- Hamiltonians of the type  \eqref{UnpertWaveHam} break isotropy in accordance with Eq. \eqref{WaveNumberNonComGeo}. This may indeed be considered a good reason to deform the Hamiltonian such that it is of the form \eqref{eqn:GUPHam}.} \\
In short, \emph{there is only one model of minimal-length quantum mechanics.}

\section{Concluding remarks}\label{sec:discussion}

While minimal-length models have been investigated for quite some time now in the context of quantum gravity phenomenology, a clear definition of what the minimal length exactly entails had not been given up until now. We have closed this gap by showing that it boils down to a cut-off in the space of wave numbers, \ie the conjugates to the positions. This cut-off is quantitatively related to the minimal length. Providing a suitable definition of wave numbers on noncommutative backgrounds, we have generalized the relation to models including coordinate noncommutativity. 

The relation between the minimal-length scale and the bound in wave-number space makes it possible to use the framework introduced here to check specific deformed Heisenberg algebras for the existence of minimal lengths.
Considering some of the most common models, we have found that one of the original ans\"atze \cite{Maggiore:1993kv}, contrary to claims in the literature, does not entail a minimal length.

A most important property of the minimal length we have distilled in this paper consists in it being solely kinematical:
every model with a bound in wave-number space contains a minimal length, independently of the Hamiltonian underlying the dynamics. Apart from that, introducing a momentum operator $\hp=\hp(\hk),$ while possibly making (especially perturbative) calculations more tractable, just amounts to a change of variables. Making the choice of Hamiltonian dependent on change of variables inherits a degree of arbitrariness. It is not a direct effect of the minimal length.

\begin{acknowledgements}
The authors acknowledge networking support by the COST Action CA18108 and  would like to thank M. Fadel and M. Maggiore for the helpful conversation. L.P. is grateful to the ``Angelo Della Riccia'' foundation for the awarded fellowship received to support the study at Universit\"at Ulm.
\end{acknowledgements}

\bibliographystyle{unsrt}
\bibliography{bib.bib}

\appendix

\section{Representation-independent proof of Eq. \eqref{BoundSeveralD}}
\label{apx:min_len_non_comm}

Let us introduce the auxiliary operators $\hat{X}_a$ such that
\begin{equation}
    [\hat{X}_a,\hat{k}_b] = i \delta_{ab}.
    \label{eqn:comm_Xk}
\end{equation}
Then, based on Eq. \eqref{eqn:comm_xk}, the position operator can be written as
\begin{equation}
    \hat{x}_a
    = \hat{X}_a + \sum_{b \neq a} g \hp_a \int_0^{\hp_b} \frac{\theta}{g^2} \D \Pi_b \hat{X}_b
    = \hat{X}_a + g \sum_{b \neq a} \frac{\partial k_b}{\partial p_a} \hat{X}_b.
\end{equation}
We are interested in studying the operator $\hat{x}_d^2$.
For this purpose, it is useful to compute the following commutator
\begin{equation}
    \left[\hat{x}_a,\hat{X}_a\right]
    = \left[\hat{X}_a + g \sum_{b \neq a} \frac{\partial k_b}{\partial p_a} \hat{X}_b , \hat{X}_a\right]
    = \left[g \sum_{b \neq a} \frac{\partial k_b}{\partial p_a} , \hat{X}_a\right] \hat{X}_b
    = - i \sum_{b \neq a} \left(\frac{\partial}{\partial k_a} g \frac{\partial k_b}{\partial p_a}\right) \hat{X}_b.
\end{equation}
Furthermore, since the commutation relation in Eq. \eqref{eqn:comm_Xk} is diagonal, we can consider a state $|\psi\rangle$ which is a common eigenstate of $\hat{X}^2_d$ with eigenvalue $\lambda$ and of $\hat{k}_b$ with eigenvalue $\bar{k}_b$, for all $b \neq d$, that is
\begin{equation}
    |\psi\rangle
    = |\lambda\rangle \bigotimes_{b \neq d} |\bar{k}_b\rangle.
\end{equation}
To further simplify the analysis below and without loss of generality, we can choose a reference frame in which all eigenvalues $\bar{k}_b = 0$.\footnote{For an argument in favor of this being the most interesting frame of reference to consider, see Section \ref{sec:noncom}.}
Notice that, since such a state is an eigenstate of $\hat{k}_b$ and since $\hat{k}_b = \hat{p}_b$ in the limit $\bar{k}_b \to 0$, we have
\begin{equation}
    \hat{p}_b |\bar{k}_b = 0\rangle = 0.
\end{equation}
Moreover, using the definition in Eq. \eqref{WaveNumberNonComGeo}, it is easy to show that
\begin{align}
    \frac{\partial \hat{k}_b}{\partial \hat{p}_a} |\bar{k}_b=0\rangle = \delta_{ab} |\bar{k}_b &= 0\rangle,&
    \frac{\partial^2 \hat{k}_b}{\partial \hat{p}_c \partial \hat{p}_a} |\bar{k}_b=0\rangle = 0.
\end{align}
However, higher order derivatives may not (and in general do not) vanish.
We then find
\begin{equation}
    \hat{x}_d |\psi\rangle
    = \left[\hat{X}_d + g \sum_{b \neq d} \frac{\partial k_b}{\partial p_d} \hat{X}_b\right]|\psi\rangle
    = \left[\hat{X}_d + \hat{X}_b g \sum_{b \neq d} \frac{\partial k_b}{\partial p_d} - i \sum_{b \neq d} \left(\frac{\partial}{\partial k_d} g \frac{\partial k_b}{\partial p_d}\right)\right]|\psi\rangle
    = \hat{X}_d |\psi\rangle,
\end{equation}
where we used the fact that
\begin{equation}
    \left(\frac{\partial}{\partial k_a} \frac{\partial k_b}{\partial p_a}\right)|\psi\rangle
    = \left(\frac{\partial p_l}{\partial k_a} \frac{\partial^2 k_b}{\partial p_a \partial p_l}\right)|\psi\rangle
    = 0.
\end{equation}
Similarly, we obtain
\begin{equation}
    \hat{x}^2_d |\psi\rangle
    = \hat{x}_d \hat{X}_d |\psi\rangle
    = \left[\hat{X}_d^2 - i \sum_{b \neq d} \left(\frac{\partial}{\partial k_d} g \frac{\partial k_b}{\partial p_d}\right) \hat{X}_b\right]|\psi\rangle.
\end{equation}
Computing the expectation value of $\hat{x}_d^2$ on the state $|\psi\rangle$ then yields
\begin{equation}
    \langle \hat{x}^2_d \rangle
    = \langle \hat{X}^2_d \rangle
    = \lambda.
\end{equation}
Furthermore, since we can always choose a reference frame in which $\langle \hat{x}_a \rangle = \langle \hat{X}_a \rangle = 0$ and since the current model does not present any preferred position, we find that, for the state $|\psi\rangle$, $(\Delta x)^2 = (\Delta X)^2 = \lambda$.
Finally, since the uncertainty relation between $X_d$ and $k_d$ is Heisenberg-like, then we can directly apply the argument used for the one-dimensional case.
We then obtain that a minimal uncertainty for $X_d$ (and therefore for $x_d$) exists if and only if the operator $\hat{k}_d$ is bounded.
Specifically,
\begin{equation}
    \Delta x_d = \sqrt{\lambda} = \frac{\pi}{2 B},
\end{equation}
where $B$ is the bound to wave-number space.

\section{Explicit proof of infinite localizability for $\alpha=1/2$}\label{app:maggiore}

We choose a series of states, $\{|\sigma\rangle\}$, with $\sigma > 0$ a parameter with units of momentum, whose normalized wave functions are given by
\begin{equation}
    \psi_\sigma(p) 
    = \langle p | \sigma \rangle
    = \frac{1}{\sqrt{\sigma \sqrt{\pi}}} e^{- \frac{k^2(p)}{2 \sigma^2}},
    \label{eqn:wavefunction}
\end{equation}
where $k(p) = \frac{\text{arcsinh}(\sqrt{\beta} p)}{\sqrt{\beta}}$ is the wave number associated with the model $\alpha=1/2$ in Eq. \eqref{eqn:modelclass}.
Then, the expectation values of the momentum and its square result as
\begin{align}
    \langle \hat{p} \rangle
    &= 0, &
    \langle \hat{p}^2 \rangle
    &= \frac{e^{\beta  \sigma^2}-1}{2 \beta }.
\end{align}
Thus, the momentum uncertainty increases with $\sigma$ as expected.
As for the expectation value of the position and its square, we obtain
\begin{align}
    \langle \hx \rangle
    &= 0, &
    \langle \hx^2 \rangle
    &= \frac{1}{2 \sigma^2}.
    \label{eqn:uncq}
\end{align}
Therefore, the uncertainty product for such states is given by
\begin{equation}
    \Delta x \Delta p 
    = \frac{e^{\frac{\beta  \sigma^2}{4}} \sqrt{\sinh(\frac{\beta \sigma^2}{2})}}{\sqrt{2 \beta} \sigma}.
\end{equation}
On the other hand, from the Robertson-Schr\"odinger relation one can straightforwardly check that
\begin{equation}
    \Delta x \Delta p
    \geq \frac{|\langle[q,p]\rangle|}{2}
    = \frac{1}{2} e^{\frac{\beta  \sigma^2}{4}}
    = \Omega,
    \label{eqn:uncrel}
\end{equation}
where $\Delta$ denotes the minimum value for the uncertainty product compatible with the model.
We observe that
\begin{equation}
    \frac{\Delta x \Delta p}{\Omega} 
    = \sqrt{2} \frac{\sqrt{\sinh\left(\frac{\beta \sigma^2}{2}\right)}}{\sqrt{\beta} \sigma} \geq 1 \qquad \forall \sigma >0,
\end{equation}
which means that the states described in Eq. \eqref{eqn:wavefunction} are compatible with the uncertainty relation implied by the model. In other words, they are part of the physical Hilbert space of the theory.
However, from Eq. \eqref{eqn:uncq}, we get
\begin{equation}
    \lim_{\sigma\to\infty} \Delta x = 0.
\end{equation}
Thus, the one-parameter family of states in Eq. \eqref{eqn:wavefunction} satisfying the uncertainty relation in Eq. \eqref{eqn:uncrel} has vanishing uncertainty in position in the limit $\sigma\to\infty$.
To put it differently, the model does not predict a minimal uncertainty in position.  
\end{document}